\documentclass[journal,twocolumn]{IEEEtran}

\usepackage{subfigure}
\usepackage{epsfig,amsmath,amssymb,epsf,cite,algorithm,algorithmic}
\usepackage{graphicx}
\usepackage{color}

\usepackage{stmaryrd}
\usepackage{amssymb}
\usepackage{tipa}
\usepackage{setspace}

\def\b0{{\pmb{0}}}

   \def\bu{{\mathbf{u}}}
   \def\bv{{\mathbf{v}}}

 \def\bH{{\mathbf{H}}}

\begin{document}

\title{A 2.0 Gb/s Throughput Decoder for QC-LDPC Convolutional Codes
\thanks{Manuscript received June 16, 2012; revised October 8, 2012. 
This paper was recommended by Associate Editor Jun Ma.}
\thanks{The work described in this paper was supported by a grant from
the RGC of the Hong Kong SAR, China (Project No. PolyU 519011).
}
\thanks{Chiu-Wing Sham, Francis C.M. Lau (corresponding author) and Wai~M. Tam are with the Department of Electronic and Information Engineering, The Hong Kong Polytechnic University, Hong Kong (email: [encwsham,encmlau]@polyu.edu.hk,tamwm@encserver.en.polyu.edu.hk).
}
\thanks{Xu Chen and Yue Zhao were with the Department of Electronic and Information Engineering, The Hong Kong Polytechnic University, Hong Kong (email: chenxugz@gmail.com,zhaoyuemagic@gmail.com). 
Xu Chen is working towards the Ph.D. degree at Northwestern University, USA and Yue Zhao is working
at the Qualcomm research center, Beijing, China.
}
\thanks{The first author and the second author have equal contributions to the paper.
}
\thanks{Digital Object Identifier xxxxxxxxx.}
\thanks{Copyright (c) 2012 IEEE. Personal use of this material is
permitted. However, permission to use this material
for any other purposes must be obtained from the IEEE by
sending an email to pubs-permissions@ieee.org.}}

\author{Chiu-Wing Sham,~\IEEEmembership{Member,~IEEE,} Xu Chen, Francis C.M. Lau,~\IEEEmembership{Senior Member,~IEEE}, Yue Zhao, \\  and
Wai M. Tam
}

\maketitle

\begin{abstract}
This paper propose a decoder architecture for low-density parity-check convolutional code (LDPCCC). Specifically, the LDPCCC is derived from a quasi-cyclic (QC) LDPC block code. By making use of the quasi-cyclic structure, the proposed LDPCCC decoder adopts a dynamic message storage in the memory and uses a simple address controller. The decoder efficiently combines the memories in the pipelining processors into a large memory block so as to take advantage of the data-width of the embedded memory in a modern field-programmable gate array (FPGA). A rate-5/6 QC-LDPCCC has been implemented on an Altera Stratix FPGA. It achieves up to 2.0 Gb/s throughput with a clock frequency of 100 MHz. Moreover, the decoder displays an excellent error performance of lower than $10^{-13}$ at a bit-energy-to-noise-power-spectral-density ratio ($E_b/N_0$) of 3.55 dB.
\end{abstract}


\section{Introduction}
\label{sec:intro}

 Low-density parity-check (LDPC) codes, first invented by Gallager in 1960's \cite{G62}, have been found to be capable of approaching the channel capacity. Later, LDPC convolutional codes (LDPCCCs) have been shown to outperform LDPC block codes in terms of error performance (e.g., lower error floors and higher coding gains) under a similar decoding complexity~\cite{FZ99}.
The comparisons between LDPCCCs and LDPC block codes from the perspectives of hardware complexity, delay requirements, memory requirements have been discussed in \cite{DPBZ09} and \cite{AEPusane08}.

LDPCCC has inherited the basic structure of convolutional code and enables a continuous encoding and decoding of messages of  varying lengths. Such a property has made LDPCCC a promising solution in many applications.
When designing an LDPCCC for an application, furthermore, many factors such as
code rate, sub-block length, coding gain, throughput, error performance
and
the encoder/decoder complexity may have to be taken into consideration.
High data rate optical communications require powerful error correction codes with low redundancies to achieve an error floor lower than a bit error rate (BER) of $10^{-13}$, preferably $10^{-15}$~\cite{Mizuochi09,Djordjevic09}. Motivated by such applications, the goal of this work is to design and implement an efficient decoder architecture such that codes can achieve \textit{high throughput}, \textit{high coding gain}, \textit{high code rate} and \textit{low error floor}.

Designing high-throughput decoder architectures for LDPC block codes has been extensively studied. In \cite{MS03},  a high-throughput memory-efficient decoder architecture   that jointly optimizes the code design, the decoding algorithm and the architecture level has been proposed.  A practical coding system design approach has been presented in \cite{ZZ05} whereby the LDPC codes are constructed subject to decoder hardware constraints. Simulation results have shown that the codes constructed suffer from only minor performance loss compared with unconstrained ones.
In \cite{WC07}, a quasi-cyclic LDPC (QC-LDPC) decoder architecture that achieves a throughput of $172$ Mbps has been studied. The high throughput is achieved by reducing the critical path through modifying the decoding algorithm as well as the check-node and variable-node processor architectures. 
In \cite{ZHW09}, the throughput of a QC-LDPC decoder is further improved by parallelizing the processing of all layers in layered decoding. Subsequently, the decoder can achieve a maximum throughput of $2.2$ Gbps with an operating frequency of $950$~MHz and $10$ min-sum decoding iterations. 
In \cite{ZW10}, the authors have proposed a high-speed flexible shift-LDPC decoder that can adapt to different code lengths and code rates. The decoder employs the Benes network to handle the complicated interconnections for various code parameters. It adopts the single-minimum min-sum decoding and achieves a throughput of $3.6$ Gbps with an operating frequency of $290$ MHz. 

Although LDPCCC decoders may ``borrow'' some design techniques used in the LDPC block decoder architectures, overall they are very different from the block code counterparts due to the distinct code construction mechanism and unique characteristics of LDPCCCs. High-throughput LDPCCC decoder architectures based on parallelization have been studied in \cite{TMKF08, MTBF07}. Such architectures can achieve a throughput of over $1$ Gbps with a clock frequency of $250$ MHz. They, however, are confined to time-invariant LDPCCCs and cannot be easily applied to time-varying ones, which usually produce a better error performance. In \cite{RSwamy07}, a register-based decoder architecture attaining up to $175$ Mbps throughput has been proposed. This architecture has successfully implemented a pipeline decoder with $10$ processing units. Nonetheless, its register-intensive architecture has limited its power efficiency.
In \cite{SBates08, SBates05}, a low-cost low-power memory-based decoder architecture that uses a single decoding processor has been proposed. On one hand, the \textit{serial} node operation uses a small portion of the field-programmable gate array~(FPGA) resources. On the other hand, such a design has posed a significant limitation on the achievable throughput. Subsequently, the memory-based designs with \textit{parallel} node operations have been proposed and have led to a substantial improvement in throughput \cite{TLBrandon09, ZChen10, CBK08}. The high throughput accomplished under these designs, however, is achieved at the cost of a complicated switch network.

To the best of the authors' knowledge, the previously proposed LDPCCC decoder architectures mainly handle random time-varying LDPCCCs.
In this paper, we propose a decoder architecture for LDPCCCs with regular structures. In particular, the proposed decoder caters for a class of LDPCCCs that have a quasi-cyclic structure and can be derived from a QC-LDPC block code \cite{F04}. The motivation of considering codes with regular structures is twofold. First, LDPCCCs with regular structures have recently attracted much interest both theoretically and empirically \cite{DAA-2011, STR-2012}. Second, following the insights from LDPC block codes, regular codes can make the decoder structure much simpler and at the same time achieve  good error performance. Therefore, developing an efficient architecture for regular codes is of high importance in practice.

The contributions in our paper are distinct from previous works in many aspects including complexity, throughput, reliability and scalability.
Firstly, we eliminate all switch networks, which are included in most of the previous implementations and are very complex for a high-rate LDPCCC.
Instead, we propose the use of dedicated block processing units, with which
we can provide higher throughput with similar decoder complexity.
Second, the quantized sum-product algorithm (QSPA) applied in our LDPCCC decoder is more reliable compared with the min-sum-based LDPCCC decoder, i.e., QSPA outperforms the min-sum-based decoder in terms of error performance. 
 Furthermore, our proposed QSPA implementation has a complexity only linearly proportional
 to the check-node degree. 
Third, it is known that more decoding iterations can enhance the error performance of the decoder.
In our decoder design, each decoding iteration is accomplished by one processor
and the processors are serially connected.
Our decoder architecture also enables us to change the number of processors easily without re-designing the whole decoder. Thus, our decoder is scalable in terms of the number of processors.
We have implemented our decoder architecture for a rate $5/6$ LDPCCC in an
Altera Stratix FPGA. The decoder has produced a throughput of $2.0$ Gbps with a clock running at $100$ MHz.
Moreover, the LDPCCC has an excellent error performance, achieving an error of lower than $10^{-13}$ at a bit-energy-to-noise-power-spectral-density ratio ($E_b/N_0$) of $3.55$ dB.

The rest of the paper is organized as follows. Section~\ref{sec:review} reviews the construction of QC-LDPCCCs and the decoding process for such codes. Section~\ref{sec:dec} describes the proposed decoder architecture and pipeline schedule. Section~\ref{sec:results} presents the implementation complexity of the decoder architecture. The FPGA simulation results are also presented in this section. Finally, Section~\ref{sec:conclude} concludes the paper.

\section{Review of LDPC Convolutional Codes}
\label{sec:review}

\begin{figure*}[t]
\begin{equation}
\begin{bmatrix}
\bH_{0}(0) & & & & & & \\
\bH_{1}(1) & \bH_{0}(1) & & & & & \\
\vdots  & \vdots  & \ddots & & & & \\
\bH_{m_s}(m_s) & \bH_{m_s-1}(m_s) & \cdots & \bH_{0}(m_s) & & & \\
 & \bH_{m_s}(m_s+1) & \bH_{m_s-1}(m_s+1) & \cdots & \bH_{0}(m_s+1) & & \\
 & & \ddots & & & \ddots & \\
 & & \bH_{m_s}(t) & \bH_{m_s-1}(t) & \cdots & \bH_{0}(t) & \\
 & & \ddots & \ddots & & & \ddots
\end{bmatrix}
\label{eq:H}
\end{equation}
\end{figure*}

\subsection{Structures of LDPCCC and QC-LDPCCC}

The parity-check matrix of an unterminated time-varying periodic LDPCCC is shown in
\eqref{eq:H}
where $m_s$ is termed as the memory of the parity-check matrix; and $\bH_{i}(t)$,  $i=0,1,\cdots, m_s$, are $(c-b) \times c$ sub-matrices with full rank. An  LDPCCC is periodic with period $T$ if $\bH_i(t) = \bH_i(t+T)$ for all $i=0,1,\cdots, m_s$. If $T=1$, the code is time-invariant; otherwise, it is time-varying. The code rate of the  LDPCCC is given by $R=b/c$. Moreover, a coded sequence $\bv_{[0,\infty]} = [\bv_0, \bv_1, \cdots,]$ with $\bv_t = [v_{t,1},v_{t,2},\cdots, v_{t,c}]$ ($t=0,1,2,\ldots$) satisfies \[ \bH_{[0,\infty  ]} \bv_{[0,\infty]}^T = 0.  \]

Given a quasi-cyclic LDPC (QC-LDPC) block code with a base matrix of size
$n_c \times n_v$ and an expansion factor of $z$\cite{Tam:2010lr}, we can construct a
QC-LDPCCC\footnote{We define a QC-LDPCCC as an LDPCCC in which
all the elements $\bH_{i}(t)$ in the parity-check matrix $\bH$ are composed of identity matrices, cyclic-right-shifted identity matrices or zero matrices.
} as follows.

\begin{enumerate}
   \item Expand the parity-check matrix of the QC-LDPC block code into a $z n_c \times z n_v$  matrix $\bH^b$.
   \item Represent the $z n_c \times z n_v$ parity-check matrix $\bH^b$ as a $M \times M$ matrix, where $M$ is the greatest common divisor of $n_c$ and $n_v$, i.e., $M = {\rm gcd}(n_c,n_v)$. Then we have
\begin{equation} \label{eq:Hb}
    \bH^b = \left[ \begin{array}{ccc}
\bH^b_{1,1} & \cdots & \bH^b_{1,M} \\
\vdots &    & \vdots   \\
\bH^b_{M,1} & \cdots & \bH^b_{M,M} \\
\end{array} \right],
\end{equation}
where $\bH^b_{i,j}$ is a $\frac{z n_c}{M} \times \frac{z n_v}{M}$ matrix, for $i,j=1,2,\cdots,M$.

\item Split $\bH^b$ into $\bH^b_l$ and $\bH^b_u$ which correspond to the lower triangular part and the strictly upper triangular part of $\bH^b$, respectively.
 $\bH^b_l$ and $\bH^b_u$ are therefore denoted, respectively, by

$
    \bH^b_l = \left[ \begin{array}{cccc}
\bH^b_{1,1} &  & &  \\
\bH^b_{2,1} & \bH^b_{2,2} & & \\
\vdots &  \vdots  & \ddots &  \\
\bH^b_{M,1} & \bH^b_{M,2} & \cdots & \bH^b_{M,M} \\
\end{array} \right]_{M \times M}
$

and

$
    \bH^b_u = \left[ \begin{array}{cccc}
\b0 & \bH^b_{1,2} & \cdots & \bH^b_{1,M}  \\
 & & \bH^b_{2,3} &  \bH^b_{2,M} \\
 &\ddots  & \ddots & \vdots \\
 &  &  & \bH^b_{M-1,M} \\
 & & & \b0 \\
\end{array} \right]_{M \times M}.
$

\item Unwrap the parity-check matrix of the block code to obtain the parity-check matrix of a QC-LDPCCC in the form of \eqref{eq:H}, i.e.,
\begin{equation} \label{eq:Hcc}
    \bH^{cc}_{[0,\infty]} = \left[ \begin{array}{cccc}
\bH^{b}_{l} &  &   & \\
\bH^{b}_{u} & \bH^{b}_{l}  & &   \\
             & \bH^{b}_{u} & \bH^{b}_{l} & \\
             &              & \ddots & \ddots\\
\end{array} \right].
\end{equation}

\end{enumerate}

\begin{figure*}
   \centering
   \includegraphics[keepaspectratio, width=1.\textwidth]{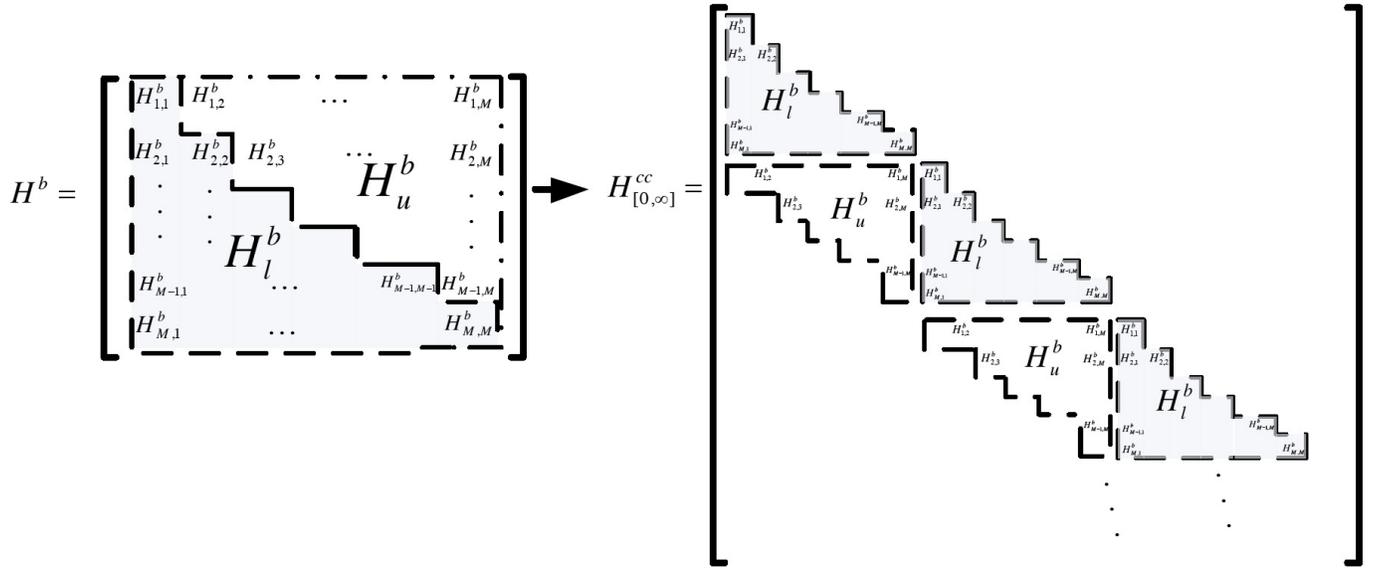}
   \caption{Illustration of constructing a QC-LDPCCC from a QC-LDPC block code.}
   \label{fig:protographLDPCCC}
\end{figure*}

The above construction process is illustrated in Fig.~\ref{fig:protographLDPCCC}. By comparing \eqref{eq:H} and \eqref{eq:Hcc}, it can be observed that the period of the QC-LDPCCC is $T = M$ and the memory $m_s$ satisfies $M= m_s+1$. It can also be observed that the relative positions between the variable nodes and the check nodes do not change. Hence the girth of the QC-LDPCCC is no less than that of the original QC-LDPC block code~\cite{CB05}. Therefore, we can construct a large-girth QC-LDPCCC by first designing the sub-matrices to obtain a large-girth QC-LDPC block code and then performing the unwrapping operation.

\subsection{Decoding Algorithm for LDPCCC}

LDPCCC has an inherent pipeline decoding process\cite{FZ99}. The pipeline decoder consists of $I$ processors, separated by $c(m_s+1)$ code symbols, with $I$ being the maximum number of decoding iterations. Throughout the decoding process, we assume that messages in log-likelihood-ratio (LLR) form are being used.

At the start of each decoding step (say at time $t_0$), the
incoming channel messages
associated with the $c$ new variable nodes $\bv_{t_0} = [v_{t_0,1}, v_{t_0,2}, \cdots, v_{t_0,c}]$ enter  the first processor.
Moreover, the corresponding variable-to-check messages for these variable nodes
have the same values as the incoming channel messages.
At the same time, the messages associated with the variable nodes $\bv_{t_0-i (m_s+1)}$
 are shifted from the $i$-th processor to the $(i+1)$-th processor, where $i=1,2,\cdots, I-1$. Then, each processor updates the $(c-b)$ check nodes
 corresponding to the $(t_0-(i-1)(m_s+1))$-th block row of $\bH_{[0,\infty]}$ in \eqref{eq:H} using
\begin{equation} \label{eq:c2vmsg}
         \alpha_{mn} = 2 \tanh^{-1}\left( \prod_{n^\prime \in \mathcal{N}(m) \backslash n} \tanh \left( \frac{\beta_{m n^\prime}}{2} \right) \right)
\end{equation}
where $\alpha_{mn}$ is the check-to-variable message from  check node $m$ to variable node $n$;  $\beta_{mn}$ is the variable-to-check message from variable node $n$ to check node $m$; $\mathcal{N}(m)$ is the set of variable nodes connected to check node $m$; and $\mathcal{N}(m) \backslash n$ is the set $\mathcal{N}(m)$ excluding  variable node $n$.
Next, the processors perform variable-node updating for
$\bv_{t_0-(i-1)(m_s+1)-m_s}$, $i=1,2,...,I,$
using
\begin{equation}
        \label{eq:v2cmsg}
        \beta_{mn} = \lambda_n + \sum\limits_{m^\prime \in \mathcal{M}(n) \backslash m} \alpha_{m^\prime n}
\end{equation}
where $\lambda_n$ is the channel message for variable node $n$;  $\mathcal{M}(n)$ is the set of check nodes connected to  variable node $n$; and $\mathcal{M}(n) \backslash m$ is the set $\mathcal{M}(n)$ excluding check node $m$.
Finally,  the \textit{a posteriori probabilities}~(APPs) for the
$c$ variable nodes $\bv_{t-(I-1)(m_s+1)-m_s}$ leaving the last processor are computed using
\begin{equation}
\beta_{n} = \lambda_n + \sum\limits_{m^\prime \in \mathcal{M}(n) } \alpha_{m^\prime n},
\end{equation}
based on which the binary value of each individual variable node is determined.

Thus, each \textit{decoding step} consists of inputting new channel messages to the decoder,
shifting messages,
updating check-to-variable messages, updating variable-to-check messages,
computing APPs and decoding the output bits.
As a result, after an initial delay of $(m_s+1)I$ decoding steps, there is a continuous
output of the decoded bits.

\section{Decoder architecture}
\label{sec:dec}

In the hardware design of an LDPCCC decoder, the processor complexity, memory requirement, throughput and error performance are closely related. It is worthwhile to study their tradeoffs so as to design a decoder meeting the application requirements. Following the notations presented in the construction of a QC-LDPCCC, we can roughly characterize the factors affecting the decoder as follows.
Suppose the decoding process is divided into $G$ stages. A smaller $G$ provides a higher level of parallelism that the decoder can achieve.  The error performance of an LDPCCC improves as $z$ increases and/or $I$ increases and/or $R$ decreases. Furthermore, the information throughput is proportional to $zR/G$ while the memory usage is proportional to $z I n_v^2 (1-R)$. Also, the processor complexity in terms of combinational logics is proportional to $z I n_v^2 (1-R)/G$. More details about the complexity of memory usage are shown in Section~\ref{sec:mem}.

It can be seen that the error performance of an LDPCCC can generally be improved at the cost of a higher processor complexity, more memory usage or a lower throughput. For instance, with the sub-matrix size $z \times z$ fixed, as the code rate $R$ decreases, the error performance becomes better at the cost of a lower information throughput. Furthermore, both the processor complexity and the memory requirement become higher due to an increase in the number of check nodes.
With the code rate and the throughput fixed, as the sub-matrix size increases, the error performance improves with the same processor complexity but more memory usage. The experiment results presented in Section~\ref{sec:results} will provide a rough guideline on how to choose the parameters in order to achieve a targeted error performance, processor complexity and memory usage.

\begin{figure}
   \centering
   \includegraphics[keepaspectratio, width=0.5\textwidth]{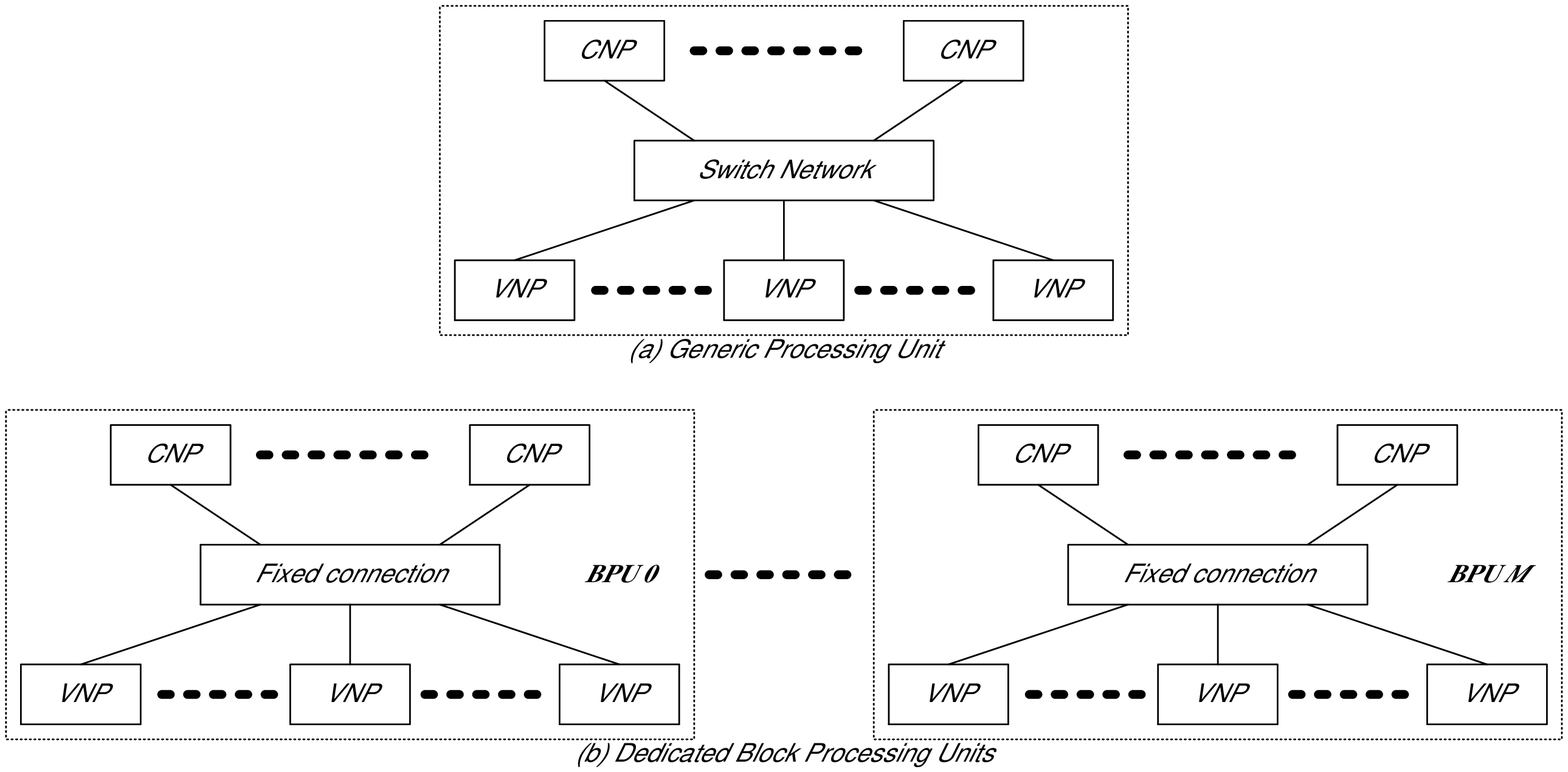}
   \caption{Generic Processing Unit and Dedicated Block Processing Unit.}
   \label{fig:BPU}
\end{figure}

In most of the previous works,  a generic processing unit such as that shown in Fig.~\ref{fig:BPU}(a) is applied in the LDPCCC decoder. For this type of design, a switch network and some corresponding control logics are required.
The complexity overhead of the switch network is not a concern in the previous works mainly because
the number of edges between the check nodes and the variable nodes is small. When the number of edges between the check nodes and the variable nodes is large, e.g., for a high-throughput and high code-rate LDPCCC, the routing and hardware complexity of the switch network becomes a critical issue.

In our proposed decoder, we use dedicated Block Processing Units (BPUs)
instead of generic processing units. Consequently, the complexity of routing and switching the messages are no longer required i.e., the complex switch network is eliminated. As shown in Fig.~\ref{fig:BPU}(b), we use
$M$ BPUs in one processor.
One BPU is used during each decoding step of one codeword and $M$  BPUs are used to facilitate the pipeline of $M$ distinct codewords simultaneously.
In general, our approach can obtain a $M$ times speed-up in throughput with the pipeline of $M$ distinct codewords.
Details will be described in Section~\ref{subsec:pipeline}.

\subsection{Architecture Design}

\begin{figure*}
   \centering
   \includegraphics[keepaspectratio, width=1.\textwidth]{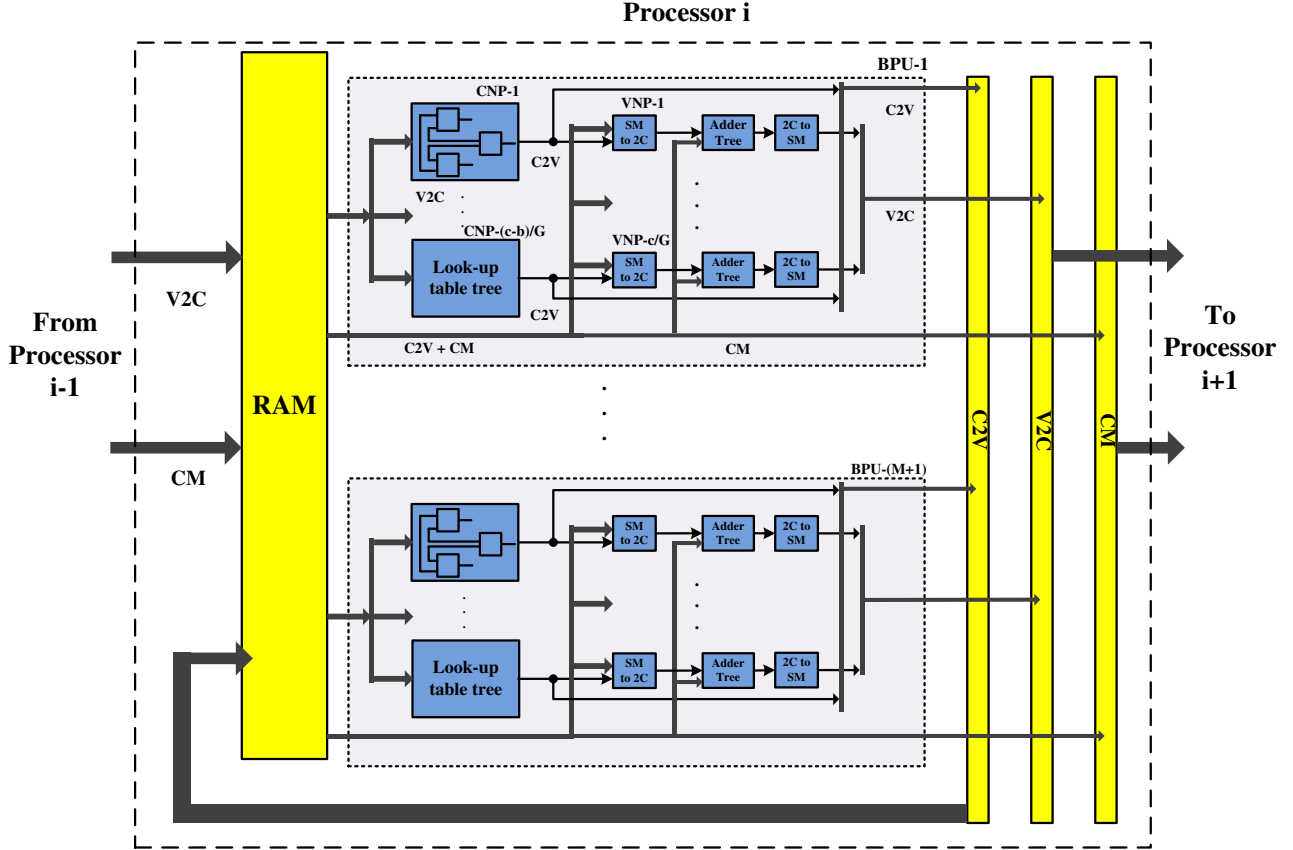}
   \caption{Block diagram of the pipeline processors in the LDPCCC decoder.}
   \label{fig:dec}
\end{figure*}

A high-throughput decoder requires parallel processing of the LDPCCC.
We propose a partially parallel decoder architecture that utilizes parallelization on both the node level and the iteration level.
 The  number of rows and the  number of columns of the sub-matrices $\bH_{i,j}^b$ in \eqref{eq:Hb} (corresponding to $\bH_{i}(t)$ in \eqref{eq:H}) are $c-b=z n_c /M$ and $c = z n_v / M$, respectively.
 Our proposed decoder architecture is illustrated in Fig.~\ref{fig:dec}. The decoder consists of $I$ processors where $I$ is the maximum number of decoding iterations. Since the memory of a QC-LDPCCC constructed using the method in Section~\ref{sec:review} is $m_s=M-1$, the variable nodes and the check nodes in each processor are separated by a maximum of $M-1$ time instants.
 Denote the $c-b$ check nodes and the $c$ variable nodes that enter a particular processor by $\bu_{t_0} = [ u_{t_0,1}, u_{t_0,2}, \cdots, u_{t_0,c-b}]$ and $\bv_{t_0} = [ v_{t_0,1}, v_{t_0,2}, \cdots,v_{t_0,c}]$, respectively.
 Then the check nodes and the variable nodes that are about to leave the processor are given by $\bu_{t_0-M+1} = [ u_{t_0-M+1,1}, u_{t_0-M+1,2}, \cdots, u_{t_0-M+1,c-b}]$ and $\bv_{t_0-M+1} = [ v_{t_0-M+1,1}, v_{t_0-M+1,2}, \cdots,v_{t_0-M+1,c}]$, respectively. At each decoding step, a BPU is responsible for processing the check nodes that enter the processor (i.e., $\bu_{t_0}$) and the variable nodes that are about to leave the processor (i.e., $\bv_{t_0-M+1}$).

At the start of each decoding step, $c-b$ check nodes are to be processed. We divide them into $G$ groups and consequently we divide a complete decoding step
into $G$ stages. At the $i$-th stage ($i=1,2,\cdots,G$), $(c-b)/G$ check nodes $[u_{t_0,(i-1)(c-b)/G+1},u_{t_0,(i-1)(c-b)/G+2}, \cdots, u_{t_0,i(c-b)/G}]$ are processed in parallel. The variable-to-check messages expressed in the sign-and-magnitude format are input to a group of $(c-b)/G$ check-node processors~(CNPs). Among the resulting check-to-variable messages, those between the check nodes in $\bu_{t_0}$ and the variable nodes \textit{not} in the set $\bv_{t_0-M+1}$ will be written to the local RAMs, waiting to be further processed by other BPUs. On the other hand, the updated check-to-variable messages between the check nodes in $\bu_{t_0}$ and the variable nodes in $\bv_{t_0-M+1}$ are converted to the format of 2's complement before being processed by the variable-node processor (VNP). Since each check node is connected to a total of $c/z$ variable nodes in $\bv_{t_0-M+1}$, $((c-b)/G)\times(c/z) = c(c-b)/Gz$ variable nodes in $\bv_{t_0-M+1}$ are connected to the newly updated check nodes and hence $c(c-b)/Gz$ VNPs are needed in one BPU. Finally, the updated variable-to-check messages are converted back to the format of sign-and-magnitude and they will be shifted to the next processor together with their associated channel messages in the next decoding step.

In the BPUs, the CNPs update the check nodes according to \eqref{eq:c2vmsg}. However, in practical implementations we need to quantize the messages to reduce the complexity. In our implementation, we adopt a four-bit quantization, where the quantization step is derived based on density evolution~\cite{CFRU01} and differential evolution~\cite{SP97}. Empirical results show that its error performance is only 0.1 dB worse than the floating-point sum-product algorithm (SPA). 

\begin{figure}[t]
   \centering
   \includegraphics[keepaspectratio, width=0.5\textwidth]{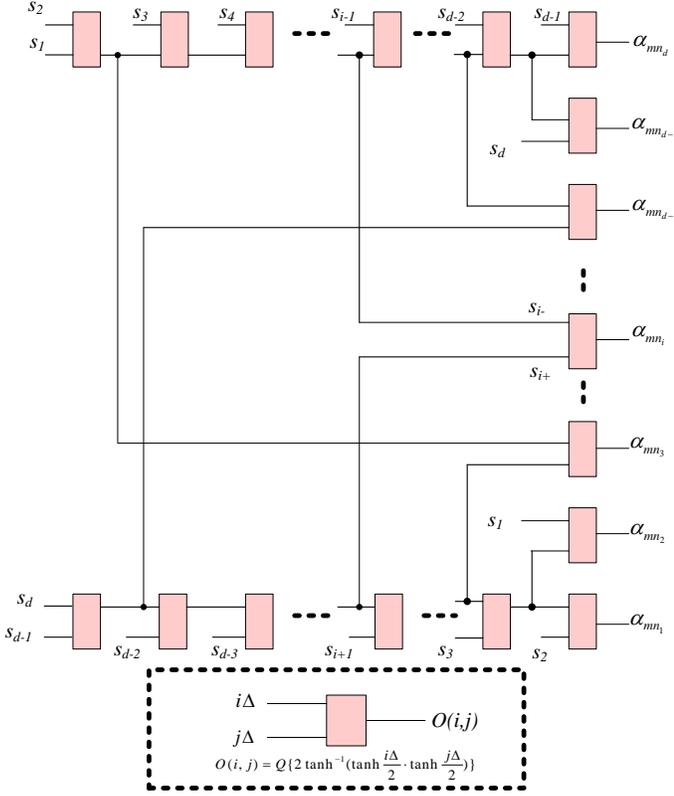}
   \caption{Implementation of a CNP using a tree of look-up tables.}
   \label{fig:cnp}
\end{figure}

We consider a check node with degree $d$. For a full quantized-SPA (QPSA) implementation, there should be $d$ inputs, each of length $4$-bits. Consequently, the size of the look-up table (LUT) becomes $2^{4d}$, which equals $2^{96}$ (as we use $d_c=24$) in our design. We can observe that it is impractical to implement such an enormous LUT. 
Here, we propose to implement the CNP with quantization (QSPA) by first pairing up the input messages and then calculating the extrinsic messages excluding the input itself. More specifically, suppose the variable nodes connected to check node $m$ is listed as $[n_1, n_2, \ldots, n_d]$ and the corresponding input messages are denoted by $[s_1,s_2,\ldots,s_d]$. The updated check-to-variable message to variable node $n_i$ is then calculated as
\begin{equation} \label{eq:cnp}
\mathcal{Q} \{ \alpha_{m n_i} \} = \mathcal{O} (s_{i-},s_{i+})
\end{equation}
where
\begin{eqnarray}
\mathcal{O}(i,j) &=& \mathcal{Q} \left\{ 2\tanh^{-1}\left( \tanh \frac{i}{2} \tanh \frac{j}{2} \right) \right\} \\
s_{i-}&=& \mathcal{O} \left( \mathcal{O} \left( \mathcal{O}(s_1,s_2) , s_3 \right), \cdots s_{i-1}\right) \\
s_{i+} &=& \mathcal{O} \left( \mathcal{O} \left( \mathcal{O}(s_d,s_{d-1}) , s_{d-2} \right), \cdots s_{i+1} \right).
\end{eqnarray}
Thus, \eqref{eq:cnp} can be implemented based on a simple LUT tree, as shown in Fig.~\ref{fig:cnp}. 
  In fact, it can be easily verified that each LUT is of size $2^{8}=256$
and the total number of units required is always $2 d=48$. Thus, 
our proposed tree-structured implementation ensures that the CNP complexity remains low, namely in $O(d_c)$.
Moreover, the VNP is basically an adding operation which can be implemented using an adder tree.

\subsection{Memory storage}
\label{sec:mem}

For clarity of presentation, we first assume $M = n_c$. Hence
we have $c-b = z$ and $c = z n_v / n_c$.
As mentioned earlier, we divide the decoding step into $G$ stages with $z/G$ check nodes being processed in parallel. We consider the $t_0$-th block row of $\bH_{[0,\infty]}^{\rm cc}$
 shown in Fig.~\ref{fig:protographLDPCCC}. This block row consists of $1 \times (n_v /n_c)$ sub-matrices, each having a size of $z \times z$.
 Thus, this block row corresponds to $z$ check nodes and $z n_v /n_c$ variable nodes in the Tanner graph.  We also assume that the $1 \times (n_v /n_c)$ sub-matrices are either the identity matrix
 or cyclic-right-shifted identity matrices. Suppose $\bu_{t_0}$ and $\bv_{t_0}$ just enter a particular processor and $\bu_{t_0-M+1}$ and $\bv_{t_0-M+1}$ are about to be shifted out of the same processor. The memory requirement is explained as follows.

\subsubsection{Storage of check-to-variable and variable-to-check messages} We denote the check nodes by $\bu_{t_0} = [u_{t_0,1},u_{t_0,2},\ldots,u_{t_0,z}]$. We further divide them into $G$ groups with the $i$-th group being denoted by $[u_{t_0,1+(i-1)z/G},u_{t_0,2+(i-1)z/G},\ldots,u_{t_0,z/G+(i-1)z/G}] \ (i =1,2,\ldots,G)$.
As explained previously, in processing $\bu_{t_0}$,
$[u_{t_0,1+(i-1)z/G},u_{t_0,2+(i-1)z/G},\ldots,u_{t_0,z/G+(i-1)z/G}]$ are processed in parallel
at the $i$-th stage
of a decoding step. Therefore in order to avoid the collisions of memory access, $z/G$ different RAMs are needed  for storing the $z/G$ messages on the edges if each of the $z/G$ check nodes is connected to only one variable node. From the construction of the QC-LDPCCC, moreover, each check node has a regular degree of $n_v$, i.e., each check node is connected to $n_v$ variable nodes. Consequently, a total of $z n_v/G$ RAMs are needed for storing the edge-messages passing between the check nodes in $\bu_{t_0}$ and their connected variable nodes to avoid the collisions of memory access. Further, each processor has $M$ sets of such check nodes, i.e., $\bu_{t_0}, \bu_{t_0-1}, \dots, \bu_{t_0-M+1}$. As a result,  $z n_v M/G$ RAMs are allocated in one processor to store the edge-messages, i.e., check-to-variable or variable-to-check messages. In addition, the data-depth and the data-width of the RAMs are equal to $G$ and the number of quantization bits, respectively.

\subsubsection{Storage of channel messages} For the channel messages, the memory storage mechanism is similar. The set of $z$ variable nodes corresponding to every $z \times z$ sub-matrix are first divided into $G$ groups.
Then $z/G$ RAMs, each of which having $G$ entries, are allocated to store the channel messages.
Moreover, the variable nodes in $\bv_{t_0}$ correspond to $n_v /n_c$  sub-matrices and each processor contains $M$ variable-node sets denoted by $\bv_{t_0}, \bv_{t_0-1}, \dots, \bv_{t_0-M+1}$. Consequently, a total of  $z n_v M/n_c G=z n_v / G$ RAMs are allocated to store the channel messages in one processor. The data-depth and the data-width of the RAMs are equal to $G$ and the number of quantization bits, respectively.

For a general case where $M$ is not necessarily equal to $n_c$,
 $z n_c n_v /G$ RAMs are needed to store the edge-messages and $z n_v M/
n_c G$ RAMs are required to store the channel messages in one processor. In modern FPGAs, the total number of internal memory bits is usually sufficient for storing the messages of codes with a reasonable length and with a reasonable number of decoding iterations.
However, the number of RAM blocks is usually insufficient. Note that the operations of the pipeline processors are identical, the connections between the RAMs and the BPUs are the same and the addresses of accessing the RAMs are the same.
By taking advantage of the homogeneity of the processors, we can combine the RAMs in different processors into one large RAM block. In particular, for the RAMs handling edge-messages, we can combine the $I$ sets of $z n_c n_v/G$ RAM blocks distributed in
the $I$ processors into \textit{one} set of $z n_c n_v/G$ RAM blocks. Similarly, for the RAMs storing the channel messages, $I$ sets of $z n_v M/ n_c G$ RAM blocks are combined into \textit{one} set of $z n_v M/ n_c G$ RAM blocks. The data-depth of the RAMs remains the same while the data-width becomes $I$ times wider. Note that the memory combination is a unique feature of LDPCCC and is not boasted by LDPC block codes\footnote{For block codes, sophisticated memory optimization has been proposed in \cite{CKLA11}. High complexity is involved and memory efficiency is achieved at the cost of a lower throughput.}.

Another advantage of such a memory storage mechanism is that the address controller is a simple counter incrementing by one at every cycle, thanks to the quasi-cyclic structure. Specifically, at the start of each decoding step, the addresses of accessing the RAMs are initialized based on the parity-check matrix $\bH_{[0,\infty]}^{cc}$. As the decoding process proceeds, the addresses are incremented by one after every stage, until all $G$ stages are completed.

\subsection{Pipeline scheduling}
\label{subsec:pipeline}

\begin{figure*}
   \centering
   \includegraphics[keepaspectratio, width=0.95\textwidth]{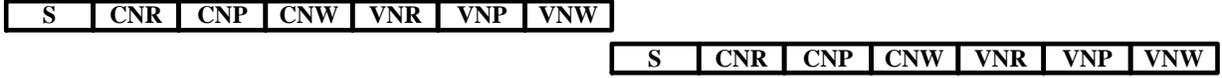}
   \caption{Conventional pipelining.
   S: Shift messages between processors;    CNR: Input messages to CN;    CNP: CN processing;
CNW: Output messages from CN;    VNR: Input messages to VN;
VNP: VN processing;    VNW: Output messages from VN.}
   \label{fig:conventional pipeline}
\end{figure*}

Conventional  LDPCCC decoder architectures \cite{MTBF07}\cite{TMKF08}\cite{RSwamy07} adopt the pipeline design shown in Fig.~\ref{fig:conventional pipeline}. Each processor sequentially does the following: shift the messages in, update the check nodes, write the data to memories, input the messages to VNP and update the variable nodes. This pipeline schedule only utilizes pipelining on the iteration level following the standard decoding process. In this paper, we propose a more efficient pipeline scheduling based on our dynamic memory storage structure.

We first describe the pipeline schedule for a single codeword. Instead of writing the updated messages from CNP and those from VNP in two separate stages, we combine them with the shifting operation. The updated messages from VNP and the channel messages associated with the updating variable nodes are directly output to the next processor, which completes the writing and shifting operations at the same time. Since some of the updated messages from CNP need not be processed by VNP, they are written to the local memories at the same time. Note that the memory locations into which the messages are shifted are exactly those storing the original messages loaded by the BPU. Therefore, there would not have any memory collisions during the process.

It can also be inferred from this process that the types of messages stored in the memories are dynamically changing.
The messages associated with $\bu_{t_0}$ are all variable-to-check messages by the time $\bu_{t_0}$ first enters a processor and is ready to be processed by CNP. After each decoding step, some of the messages are substituted by the updated variable-to-check messages from the previous processor. When $M$ decoding steps are completed, all the check-to-variable messages originally associated with $\bu_{t_0}$ will be completely substituted  by variable-to-check messages. Yet, they are now messages for $\bu_{t_0+M+1}$ and are ready for CNP in a new round of decoding.

\begin{figure*}[ht]
\centering
\subfigure[Single-codeword pipeline.]{
\includegraphics[keepaspectratio, width=1.0\textwidth]{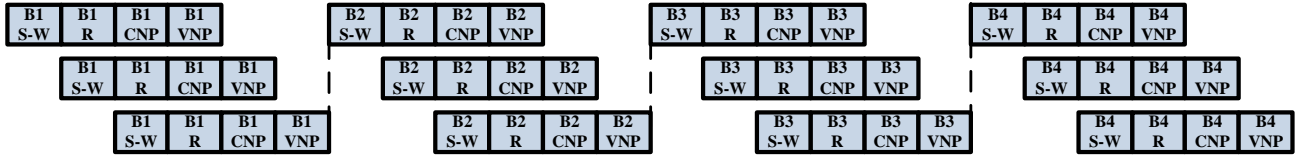}
\label{fig:pipeline2}
}
\subfigure[Multiple-codeword pipeline.]{
\includegraphics[keepaspectratio, width=1.0\textwidth]{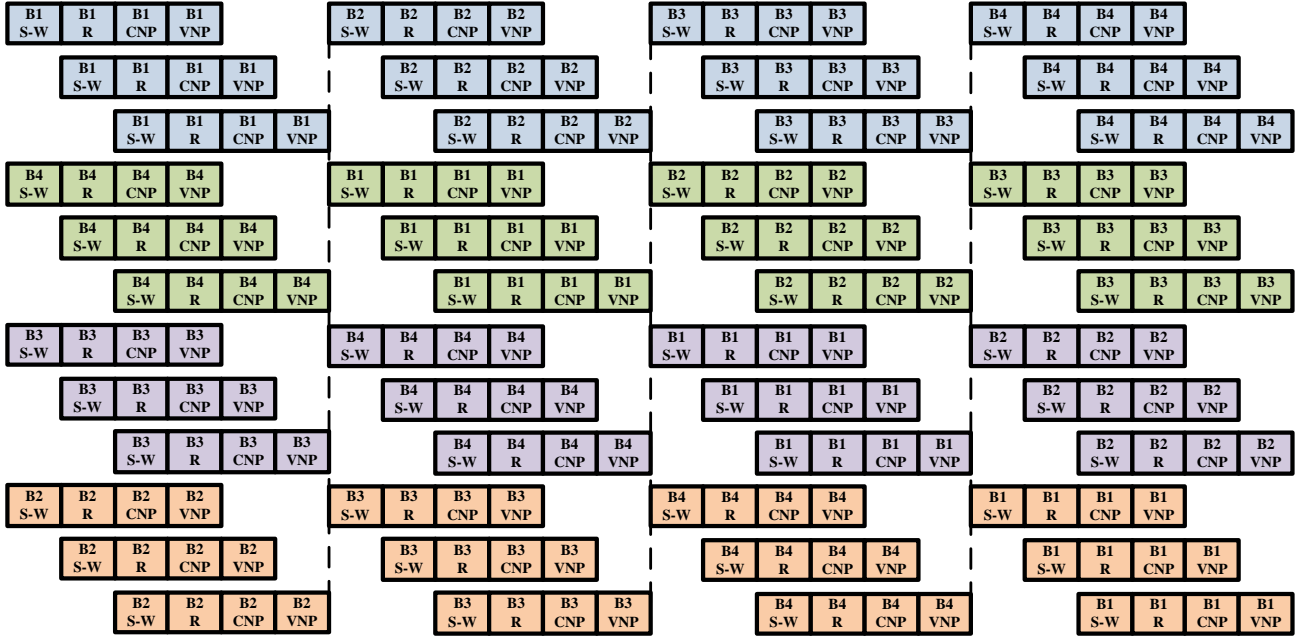}
\label{fig:pipeline3}
}
\label{fig:proposedpiepline}
\caption{Proposed Pipeline. B$i$: processing of block $i$; S-W: Shift messages and write messages to the next processor; R: Input messages to the block processing unit;
CNP: check-node processing; VNP: variable-node processing.}
\end{figure*}

Figure~\ref{fig:pipeline2} describes the pipeline for a single codeword assuming $G=3$ and $M=4$. Comparing Fig.~\ref{fig:conventional pipeline} and Fig.~\ref{fig:pipeline2}, it can be observed that decoding a group of check nodes using the proposed pipeline scheduling only takes $4/7$ of the time cost in conventional scheduling. The homogeneity of the pipeline processors also facilitates a pipeline processing of multiple codewords. As shown in Fig.~\ref{fig:pipeline2} where a single codeword is being decoded,
the processing time of different BPUs are separated in the sense that while one BPU is processing, the other BPUs remain idle. To further increase the throughput, we can schedule  other BPUs to process other codewords. Since the total number of blocks in a processor is $M$, we can incorporate a maximum of $M$ different codewords in one processor, i.e.,
allowing ${\rm BPU}_i$ to process Codeword-$i$, for $i=1,2,\cdots, M$. Depending on the number of codewords incorporated, the throughput can be increased by a factor of $M$ at the cost of additional memory storage and additional hardware complexity of the BPUs. Figure~\ref{fig:pipeline3} illustrates the pipeline schedule for four codewords with $G=3$ and $M=4$.

Using our proposed pipeline schedule, the throughput of the decoder is $(n_v - n_c) z /M$ information bits for every $G+d$ cycles, where $d$ is the time delay for each pipeline stage such that $G+d$ cycles are used by one BPU. As there are more decoding stages, i.e., $G$ increases, the throughput tends to $(n_v-n_c) z f /MG$ bits/s with a running clock of $f$ Hz.

\textit{An illustrative example of the RAM storage and decoding process}

\begin{figure*}
   \centering
   \includegraphics[keepaspectratio, width=1.05\textwidth]{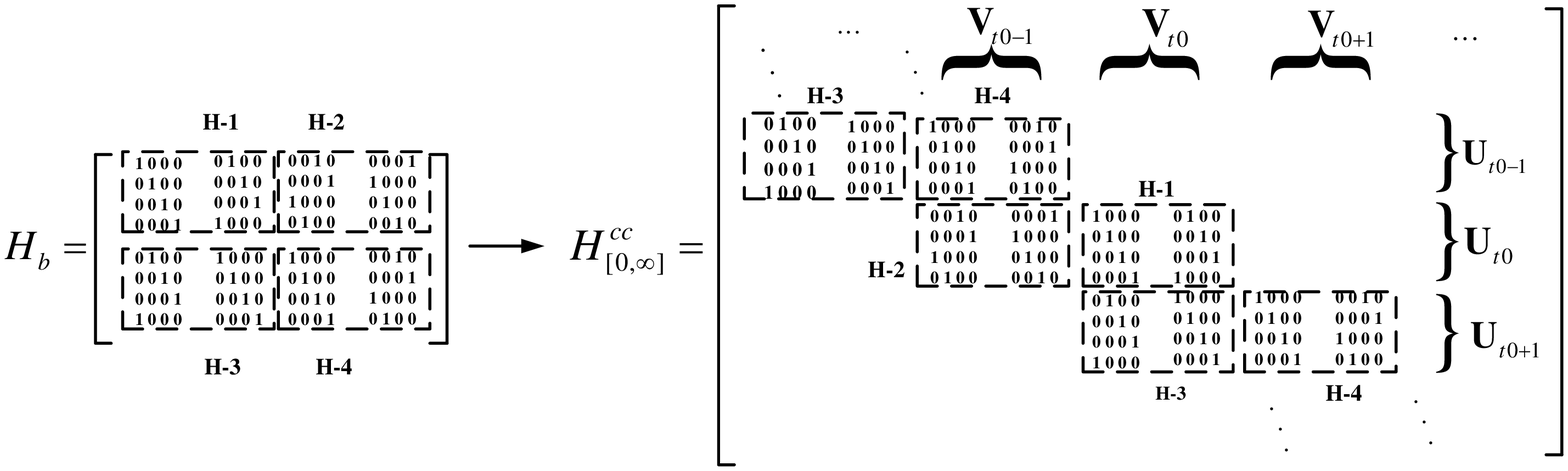}
   \includegraphics[keepaspectratio, width=1.05\textwidth]{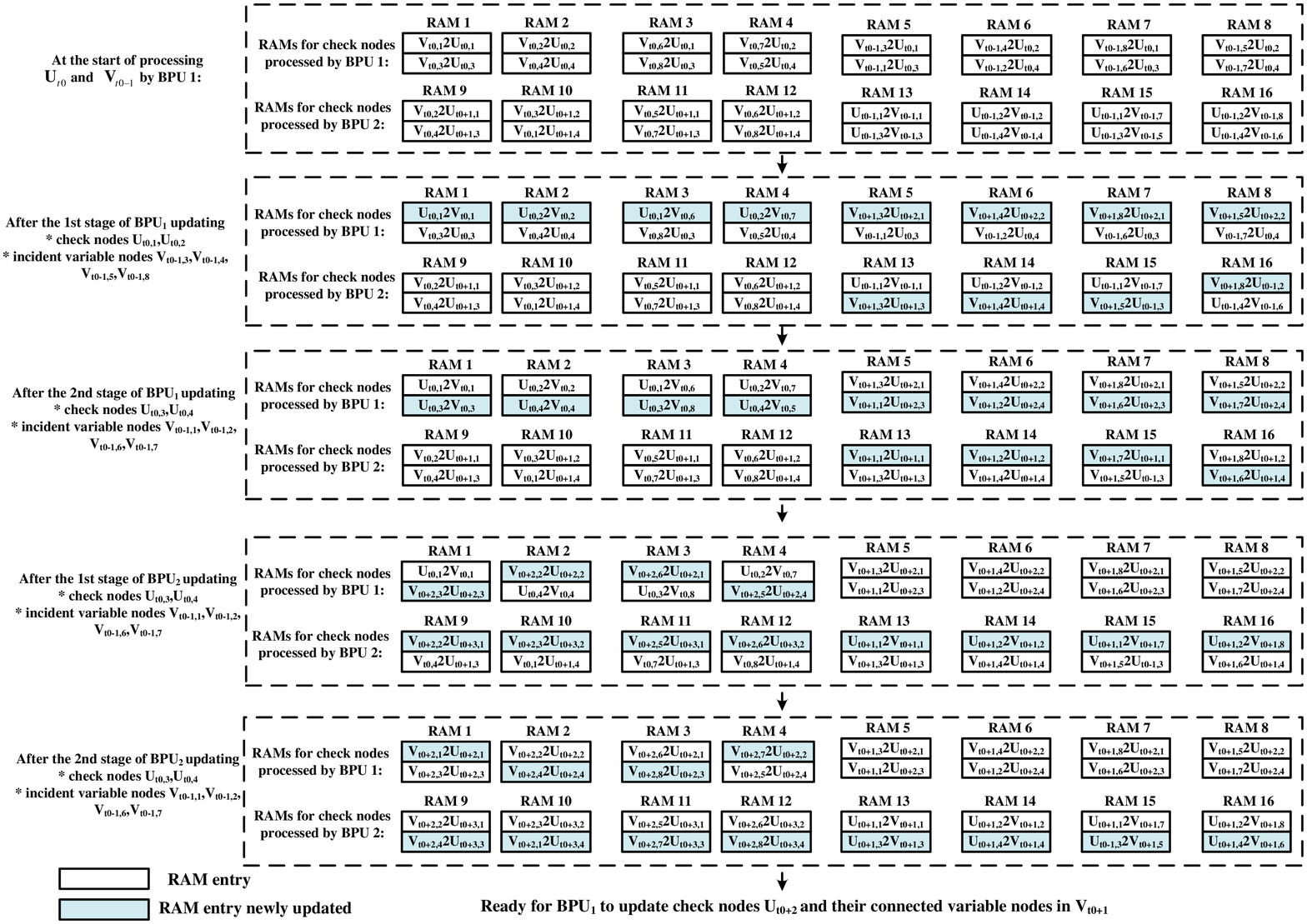}
   \caption{Example of RAM storage. $z=4$ and $G=2$.}
   \label{fig:ram example}
\end{figure*}

\textit{Example:} we consider a QC-LDPCCC with $G=2$,  $z=4$, $n_c = 2$ and $n_v =4$. Since $M ={\rm gcd}(n_c , n_v) = 2$, each processor has $M=2$ BPUs. In each processor, $z n_c n_v /M G = 8$ RAMs are dedicated to store edge-messages and $z n_v / n_c G =4$ RAMs are dedicated to store channel messages. Assume that the check nodes $\bu_{t_0} = [u_{t_0,1},u_{t_0,2},\ldots,u_{t_0,4}]$ just enter a processor and the
variable nodes $\bv_{t_0-1} = [v_{t_0-1,1},v_{t_0-1,2},\ldots,v_{t_0-1,8}]$ are about to leave. The decoding step of processing ${\rm BPU}_i$ $(i=1,2)$ is divided into $G=2$ stages. Figure~\ref{fig:ram example} shows the dynamic storage of the edge-messages in the RAMs at different time instances.

Step 1) It shows the RAM storage at the start of processing $\bu_{t_0}$ and $\bv_{t_0-1}$ by ${\rm BPU}_1$. It can be seen that RAM 1 to 8 store the variable-to-check messages for $\bu_{t_0}$ which is ready to be processed. RAM 13 to 16 store the
latest check-to-variable messages  for $\bu_{t_0-1}$, which are updated in the previous decoding step by ${\rm BPU}_2$. RAM 9 to 12 store the variable-to-check messages that are newly updated in the previous decoding step and are shifted from the previous processor.

Step 2) It shows the RAM storage after the first stage of ${\rm BPU}_1$ processing. At the first stage, ${\rm BPU}_1$ will process $u_{t_0,1}$ and $u_{t_0,2}$ and their connected variable nodes in $\bv_{t_0-1}$, e.g., $[v_{t_0-1,3},v_{t_0-1,4},v_{t_0-1,5},v_{t_0-1,8}]$. CNP reads the variable-to-check messages from the first set of entries located in RAM 1 to 8. The newly updated check-to-variable messages between $\bu_{t_0}$ and $\bv_{t_0}$ from CNP are input to the first set of entries in RAM 1 to 4  (i.e., from where the check-to-variable messages are read), while the newly updated check-to-variable messages between $\bu_{t_0}$ and $\bv_{t_0-1}$ are input to the VNP and the resulting variable-to-check messages are shifted to the next processor. As a result, the updated variable-to-check messages between $\bv_{t_0+1}$ and $\bu_{t_0+2}$ are written to RAM 5 to 8 and those between $\bv_{t_0+1}$ and $\bu_{t_0+1}$ are written to RAM 13 to 16.

Step 3) It shows the RAMs after the second stage of ${\rm BPU}_1$ processing. At the second stage, ${\rm BPU}_1$ will process $u_{t_0,3}$ and $u_{t_0,4}$ and their connected variable nodes in $\bv_{t_0-1}$, e.g., $[v_{t_0-1,1},v_{t_0-1,2},v_{t_0-1,6},v_{t_0-1,7}]$. CNP reads the variable-to-check messages from the second set of entries located in RAM 1 to 8. The newly updated check-to-variable messages between $\bu_{t_0}$ and $\bv_{t_0}$ from CNP are input to the second set of entries in RAM 1 to 4 (i.e., from where the check-to-variable messages are read), while the newly updated check-to-variable messages between $\bu_{t_0}$ and $\bv_{t_0-1}$ are input to the VNP and the resulting variable-to-check messages are shifted to the next processor. As a result, the updated variable-to-check messages between $\bv_{t_0+1}$ and $\bu_{t_0+2}$ are written to RAM 5 to 8 and those between $\bv_{t_0+1}$ and $\bu_{t_0+1}$ are written to RAM 13 to 16.

The RAM updating at the decoding step of ${\rm BPU}_2$ is analogous to
Steps 2) and 3) above. After the second stage of ${\rm BPU}_2$, RAM 1 to 8 will have the variable-to-check messages ready for $\bu_{t_0+2}$ and their connected variable nodes in $\bv_{t_0+1}$. The RAM storage is similar to that in Step 1) with the time instances incrementing by $M=2$. A new round of ${\rm BPU}_1$ updating will follow according to Steps 2) and 3).

Also note that once the address controller is initialized at the start of the $G$ stages, the read/write address of accessing the RAMs are simply incremented by 1.

\section{Experimental Results}
\label{sec:results}

We have implemented the QC-LDPCCC decoder on Altera Stratix IV. 
All the BER results for the QC-LDPCCC decoder are hence obtained from FPGA experiments under additive white Gaussian noise (AWGN) channels and $4$-bit quantization.
Based on a QC-LDPC block code with a  $4 \times 24$ base matrix,
we construct QC-LDPCCCs of different sub-matrix sizes.
Moreover, the sub-matrices of the block code are chosen such that the girth equals $8$.
Then we simulate the BER performance of the QC-LDPCCCs
under different decoding iteration numbers.
Specifically, we have implemented LDPCCC decoders with the following parameters:
(a) $z=422$ and $I=18$;  (b) $z=512$ and $I=18$; (c) $z=1024$ and $I=12$;
(c) $z=1024$ and $I=10$. Recall that $z \times z$ represents the sub-matrix size of
each entry in the $4 \times 24$ base matrix while $I$ denotes the number
  of iterations (i.e., processors) used in the LDPCCC decoders.

\begin{table*}[ht]
\begin{center}
\caption{Implementation complexity for QC-LDPCCC of different sub-matrix sizes. Code 1-S: $z= 422$, $I=18$, single-codeword. Code 2-S: $z= 512$, $I=18$, single-codeword. Code 3-S: $z=1024$, $I=12$, single-codeword. Code 4-S: $z=1024$,$I=10$, single-codeword. Code 1-P: $z= 422$, $I=18$, four-codeword pipeline. Code 2-P: $z= 512$, $I=18$, four-codeword pipeline. Code 3-P: $z=1024$, $I=12$, four-codeword pipeline. Code 4-P: $z=1024$, $I=10$, four-codeword pipeline. 
The implementation complexity of the QC-LDPC block decoder in \cite{WC07} is shown for comparison.}
\label{tab:implementation}
\small

\begin{tabular}{|l|l|l|l|l|l|l|l|l|cccccccc} \hline
& Stage No. &  Memory  & Combinational & Registers & Memory & Clock  & Throughput & Required $E_b /N_0$\\
& $G$        & depth   & ALUTs         &           & bits    &  frequency & (info bits) & at a BER of $10^{-10}$ \\ \hline
Code 1-S & $422$ &  $512$ & $106288$ & $68609$ & $4402268$ & $100$ MHz & $0.5$ Gbps & $3.42$ dB \\ \hline
Code 2-S & $512$ &  $512$ & $104938$ & $68634$ & $4402268$ &  $100$ MHz & $0.5$ Gbps & $3.40$ dB \\ \hline
Code 3-S & $1024$ &  $1024$ & $73066$ & $50087$ & $5829352$ & $100$ MHz & $0.5$ Gbps & $3.48$ dB \\ \hline
Code 4-S & $1024$ &  $1024$ & $62823$ & $43745$ & $4844140$ & $100$ MHz &  $0.5$ Gbps & $3.60$ dB \\  \hline
Code 1-P & $422$ & $512$ & $175420$ & $105427$ & $17558528$ & $100$ MHz & $2.0$ Gbps & $3.42$ dB \\  \hline
Code 2-P & $512$ & $512$ & $170102$ & $105505$ & $17558528$ & $100$ MHz &  $2.0$ Gbps & $3.40$ dB \\  \hline
Code 3-P & $1024$ & $1024$ & $134102$ & $86654$ & $23283712$ & $100$ MHz &  $2.0$ Gbps & $3.48$ dB \\  \hline
Code 4-P & $1024$ & $1024$ & $120804$ & $80342$ & $19348480$ & $100$ MHz &  $2.0$ Gbps &  $3.60$ dB \\  \hline
 Wang \cite{WC07} & --- & ---  &   $28229$ &  $26926$  &  $5800000$ &   $190$ MHz &   $0.2$ Gbps &  $4.40$ dB \\ \hline
\end{tabular}
\end{center}
\end{table*}
\normalsize

\begin{figure}[t]
   \centering
   \includegraphics[keepaspectratio, width=0.5\textwidth]{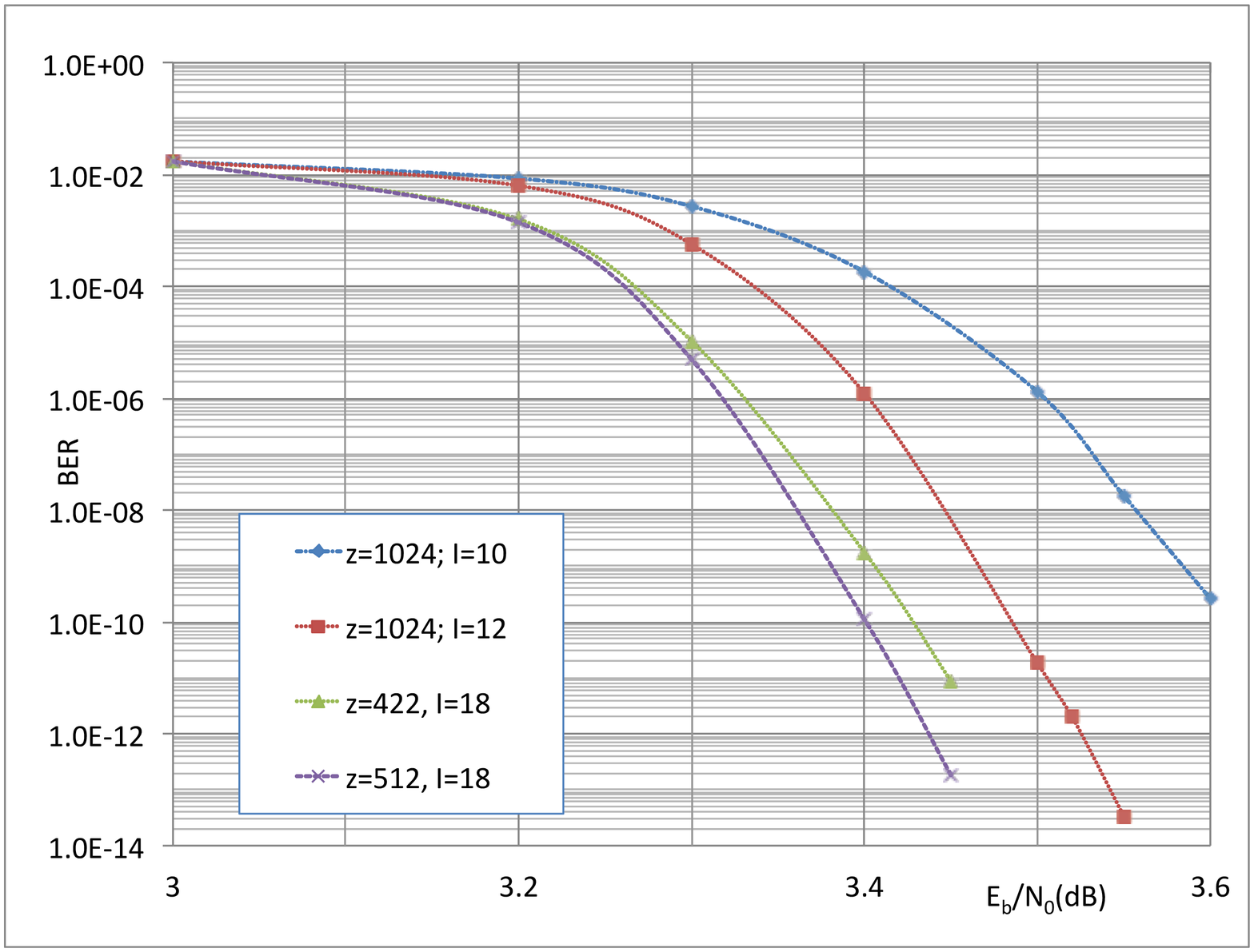}
   \caption{ Bit-error-rate (BER) results for the LDPCCCs with different sizes. 
   The results are obtained from FPGA experiments under AWGN channels and $4$-bit quantization.}
   \label{fig:ber}
\end{figure}

Table~\ref{tab:implementation} shows the hardware complexity of the decoders
when combined with the noise generator. The complexities for a single-codeword implementation as well as a four-codeword pipeline implementation are shown.
We observe that the hardware complexity increases as the code length and the number of processors increases.
Figure~\ref{fig:ber} further shows the BER results for the LDPCCCs.

Based on Fig.~\ref{fig:ber} and Table~\ref{tab:implementation}, we can see a tradeoff between (i) the BER performance, (ii)  the code length and (iii) the number of processors (i.e., the number of iterations).
We compare the performance of LDPCCC with $z=1024$ but
with different number of decoding iterations $I$.
We can see that the LDPCCC with $I=12$ is more than $0.1$~dB better than that with $I=10$ at a BER of $3 \times 10^{-10}$.
We further compare the error performance of
codes with similar processor complexity. We observe from Table~\ref{tab:implementation} that the LDPCCC using $z=1024$ and $I=12$ has a similar complexity with the ones using (i) $z=422$ and $I=18$ or (ii) $z=512$ and $I=18$.
Figure~\ref{fig:ber} shows that the LDPCCC using $z=1024$ and $I=12$ is outperformed by the ones using (i) $z=422$ and $I=18$ or (ii) $z=512$ and $I=18$, even though the latter two codes have smaller sub-matrix sizes. It is therefore obvious that a larger number of decoding iterations can help reducing the error rate even when a smaller sub-matrix size is used.
In summary, we find that the number of decoding iterations plays an important role in the error performance of the LDPCCC.

Based on the above results, the following guidelines can be used in designing a LDPCCC decoder.
\begin{itemize}
\item To increase the decoder throughput while maintaining a similar BER performance and the same number of memory bits, we can reduce the memory depth $G$ at the cost of more combinational logics.
\item To reduce the cost of combinational logics while maintaining a similar BER performance and throughput, we can increase $z$ and use a smaller number of processors $I$. Under such circumstances, the total memory bits may increase.
\item To reduce the memory bits while maintaining a similar BER performance and throughput, we can use a smaller $z$ and a larger $I$ at the cost of combinational logics.
\end{itemize}

 In addition, we attempt to compare our implementation results with those 
found from the literature.
Since the objective of our work is to achieve high throughput and good error performance, the code length and code rate of the codes used in our experiments are relatively large. 
While we can find quite a number of decoders in the literature, 
none of them consider codes with length comparable to the ones we use. All of them 
assume lengths which are relatively short and consequently they have  high error floors and small coding gains.
The ``closest'' one we can find is the QC-LDPC block decoder described by Wang and Cui \cite{WC07}, who target a high-speed decoder and adopt a length-$8176$ QC-LDPC code in the experiment.
In Table I,  we add the implementation results of the decoder in \cite{WC07}. Although the decoder in \cite{WC07} seems to be less complex than our designs, its throughput ($0.2$ Gbps) is only $1/10$ of ours ($2$ Gbps). If $10$ decoders in \cite{WC07} are put together in order to achieve the same throughput as our decoders, the total complexity of 
the decoders will become larger than ours. Furthermore, the decoder in  \cite{WC07} displays an error floor at a BER of $10^{-10}$ while our decoder does not. In fact, at a BER of $10^{-10}$, our decoders can achieve an extra coding gain of $0.8$ dB to $1$ dB over the decoder in  \cite{WC07}. Thus, our proposed decoder is superior in achieving high throughput, high coding gain and low error floor.

We also compare the BER performance of LDPCCCs and their block-code counterparts under similar processor complexity and throughput.
Compared with a single-processor decoder of an LDPC block code with the same iteration number $I$, the LDPCCC decoder with $I$ processors, the length of the coded bits stored in each processor being the code length of the block code, incurs $I$ times more complexity, but achieves $I$ times higher throughput. In order for the LDPC block decoder to attain the same throughput, $I$ times more processors are needed to decode in parallel. Under such circumstances, the overall complexity of the LDPC block decoder will increase by $I$ times and becomes the same as the LDPCCC counterpart. Therefore, the fairness of comparing LDPCCC with its block-code counter part based on which the LDPCCC is derived is validated from the perspective of processor complexity and throughput.

\begin{figure}
   \centering
   \includegraphics[keepaspectratio, width=0.5\textwidth]{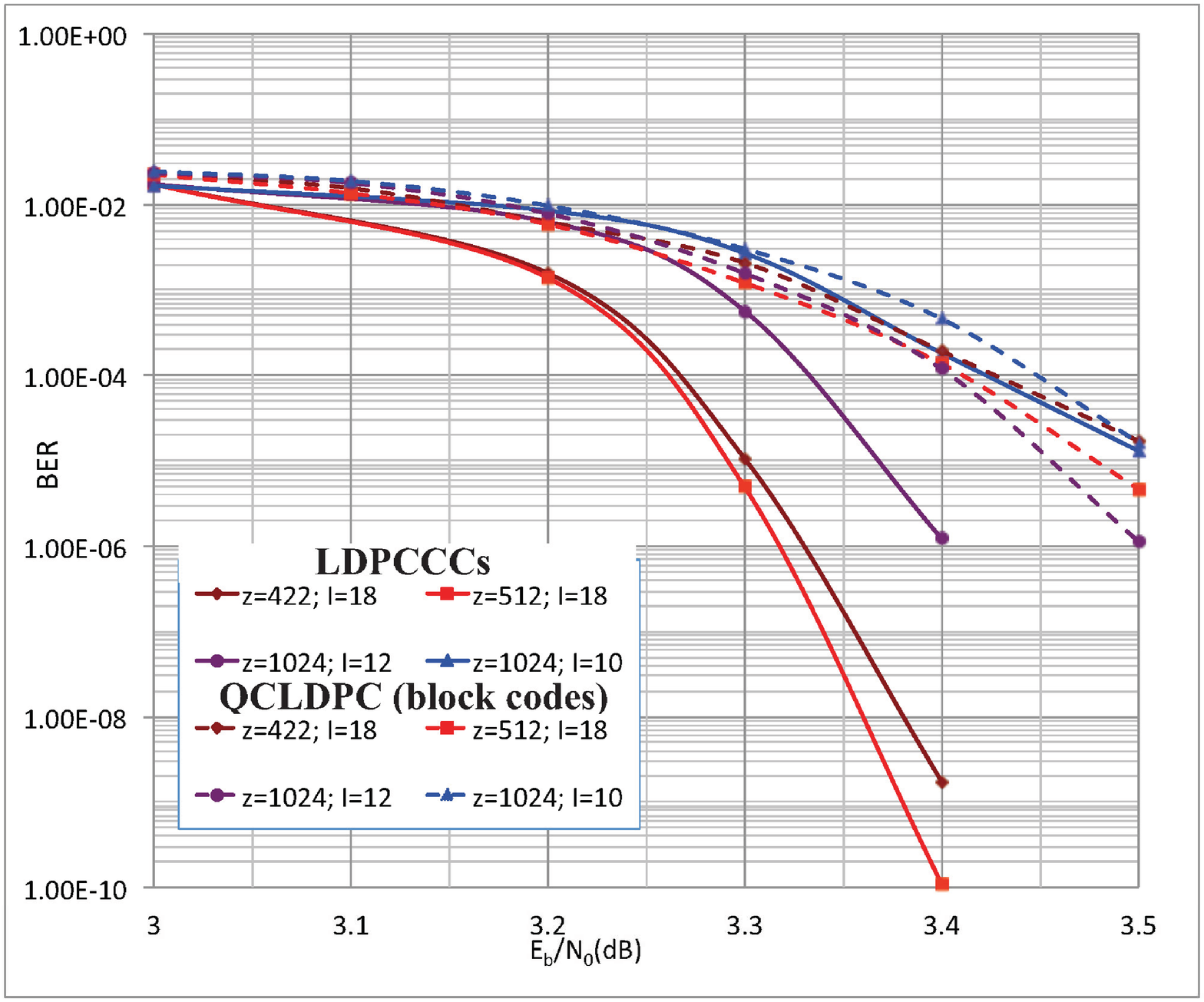}
   \caption{ Comparison of BER results between LDPCCCs and LDPC block-code counterparts under AWGN channels.
The results of the LDPCCCs and the LDPC block codes are represented by solid lines and dashed lines, respectively. The results of the LDPCCCs are obtained from FPGA experiments under $4$-bit quantization while and those of the LDPC block codes are obtained from computer simulations (using C programming) based on 4-bit quantized messages.}
   \label{fig:berqcldpcvsldpccc}
\end{figure}

Figure~\ref{fig:berqcldpcvsldpccc} shows the BER performance of LDPCCCs and their block-code counterparts. 
 The results of the LDPC block codes are obtained from computer simulations (using C programming) based on 4-bit quantized messages.
It can be seen that the BER performance of LDPCCCs are generally superior. For instance, the LDPCCC with $z=422$ and $I=18$ has a gain of $0.2$~dB at  a BER of $2 \times 10^{-5}$ over its block-code counterpart.
Another observation is that the advantage of LDPCCC over its block-code counterpart becomes obvious as the number of decoding iterations increases.
For example,
the performance of LDPCCC with $z=1024$ and $I=10$ has a similar performance of its
block-code counterpart at a BER of $2 \times 10^{-5}$; and it outperforms its
block-code counterpart by $0.1$~dB at a BER of $\times 10^{-6}$ when the number of decoding iterations increases to $12$, i.e., $I=12$.
As a result, when the number of decoding iterations is large, LDPCCC is considered to be a better choice in terms of error performance.

\section*{Acknowledgements}
The authors would like to thank the associate editor and the anonymous reviewers for their invaluable comments on the earlier version of this paper.

\section{Conclusion}
\label{sec:conclude}

An efficient partially parallel decoder architecture for QC-LDPCCC has been proposed in this paper. The dedicated Block Processing Unit is also proposed such that the complexity overhead of the switch network can be removed. Rate-$5/6$ LDPCCC decoders of different sub-matrix sizes have been implemented on an Altera FPGA with our proposed architecture.
It is found that our decoders can achieve a throughput of $2.0$~Gb/s. Experimental results further show that QC-LDPCCCs outperform their block-code counterparts under the same throughput and similar overall decoder complexity.
Moreover, the QC-LDPCCCs derived from well-designed block codes can achieve an
error floor of lower than $10^{-13}$.



\begin{IEEEbiography}
[{\includegraphics[width=1in,height=1.25in,clip,keepaspectratio]{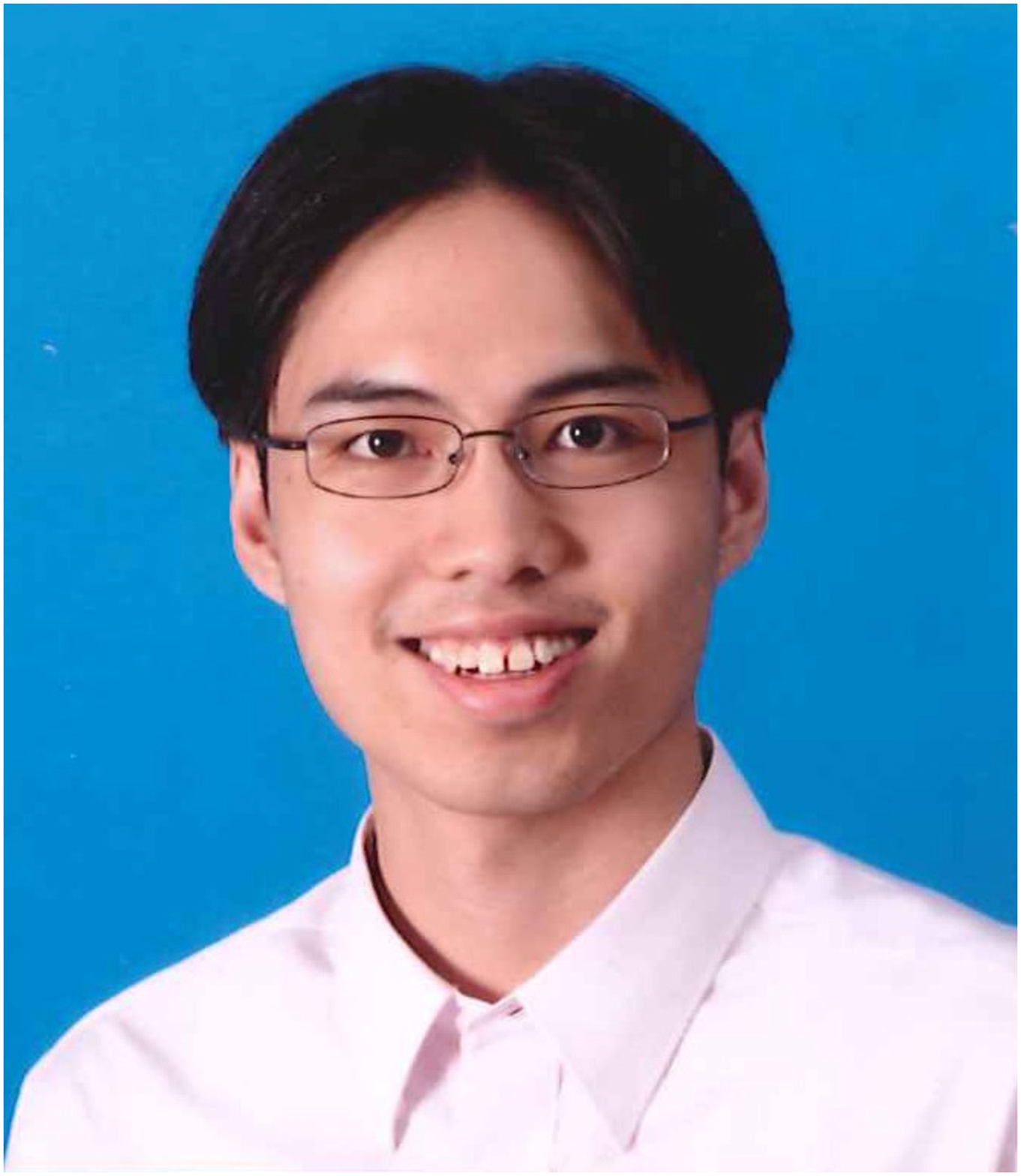}}]
{Chiu-Wing Sham} received the Bachelor degree in computer engineering, and the M.Phil. degree and the Ph.D. degree from The Chinese University of Hong Kong, Hong Kong, in 2000, 2002, and 2006, respectively. He was a Research Engineer with Synopsys, Shanghai, China, and an Electronic Engineer working on the FPGA applications of motion control system with ASM (HK). He joined the Electronic and Information Engineering Department of The Hong Kong Polytechnic University, as a Lecturer in August 2006. His research interests include design automation of VLSI, design optimization of digital VLSI systems and embedded systems.
\end{IEEEbiography}

\begin{IEEEbiography}
[{\includegraphics[width=1in,height=1.25in,clip,keepaspectratio]{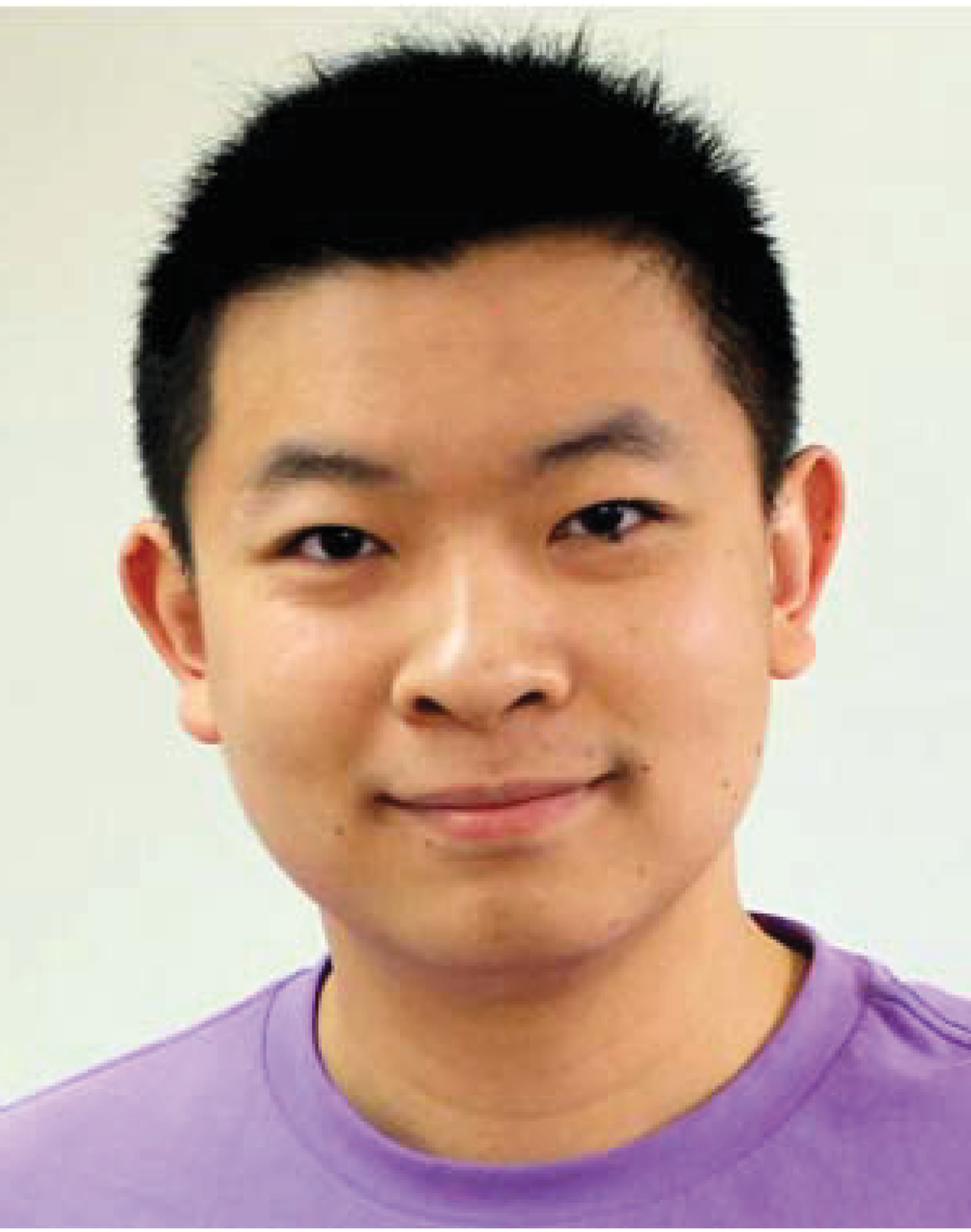}}]
{Xu Chen} received his B.E. degree from Sun Yat-sen (Zhongshan) University, China in 2007 and his M.S. degree from Purdue University, West Lafayette in 2009. From 2009 to 2011, he was a research assistant in the Hong Kong Polytechnic University, Hong Kong. He is currently working towards the Ph.D. degree at Northwestern University, USA. His research interests include coding theory, optimization and cooperative communications.
\end{IEEEbiography}

\begin{IEEEbiography}
[{\includegraphics[width=1in,height=1.25in,clip,keepaspectratio]{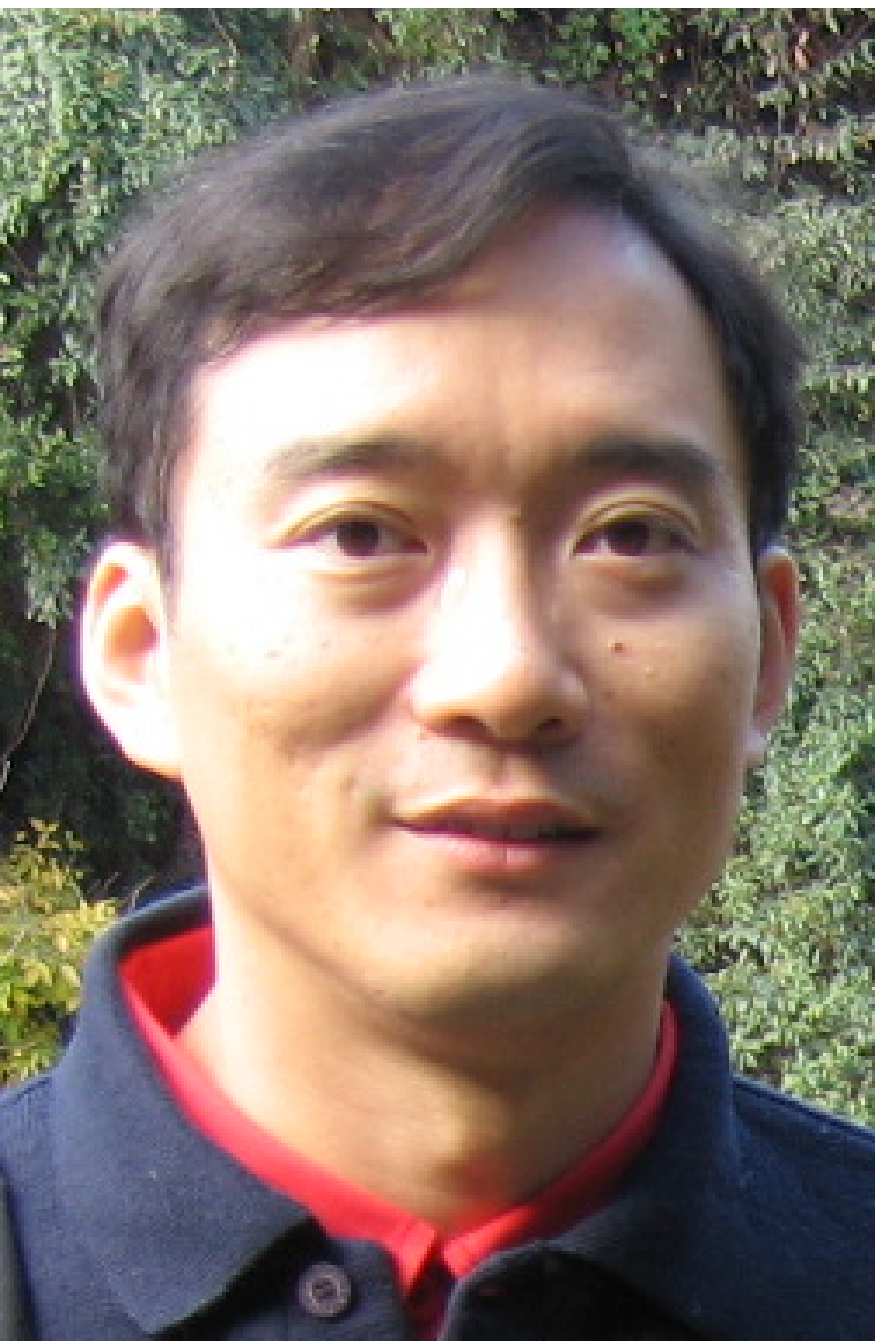}}]
{Francis C.M.~Lau} (M'93--SM'03) received the BEng~(Hons) degree
in electrical and electronic engineering and the PhD degree
from King's College London, University of London, UK, in 1989 and 1993,
respectively.\\
\hspace*{3ex}
He is a Professor and Associate Head at the Department of Electronic and Information Engineering, The Hong Kong Polytechnic University, Hong Kong.
He is also a senior member of {IEEE}. He is the co-author of {\em Chaos-Based Digital
Communication Systems} (Heidelberg: Springer-Verlag, 2003) and {\em Digital Communications with Chaos: Multiple Access Techniques and Performance Evaluation}
(Oxford: Elsevier, 2007). He is also a co-holder of three US patents and one pending US patent. He has published over 230 papers. His main research interests include channel coding, cooperative networks, wireless sensor networks, chaos-based digital communications, applications of complex-network theories, and wireless communications.\\
\hspace*{3ex}
He served as an associate editor for {\em IEEE Transactions
on Circuits and Systems II} in 2004--2005 and  {\em IEEE Transactions
on Circuits and Systems I} in 2006--2007. He was also an associate editor of {\em Dynamics of
Continuous, Discrete and Impulsive Systems, Series B} from 2004 to 2007, a co-guest editor of {\em Circuits,
Systems and Signal Processing} for the special issue ``Applications of Chaos in Communications'' in 2005, and an associate editor for IEICE Transactions (Special Section on Recent Progress in Nonlinear Theory and Its Applications) in 2011. He has been a guest associate editor of  {\em International Journal and Bifurcation and Chaos} since 2010 and an associate editor of {\em IEEE Circuits and Systems Magazine} since 2012.
\end{IEEEbiography}

\begin{IEEEbiography}
[{\includegraphics[width=1in,height=1.25in,clip,keepaspectratio]{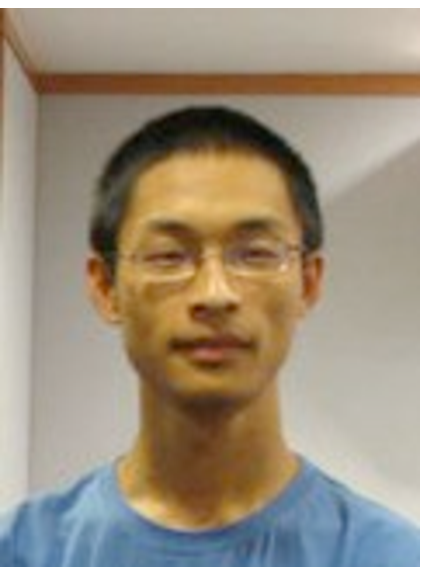}}]
{Yue Zhao} Yue Zhao received the BE degree in information Engineering
from Shanghai Jiaotong University, China in 2009. He was a postgraduate
student and research assistant at the Hong Kong Polytechnic
University, Hong Kong, from 2009 to 2012, where he was working on algorithms
and implementations for the LDPC decoding. He is currently working
at the Qualcomm research center, Beijing, China.
\end{IEEEbiography}

\begin{IEEEbiography}
[{\includegraphics[width=1in,height=1.25in,clip,keepaspectratio]{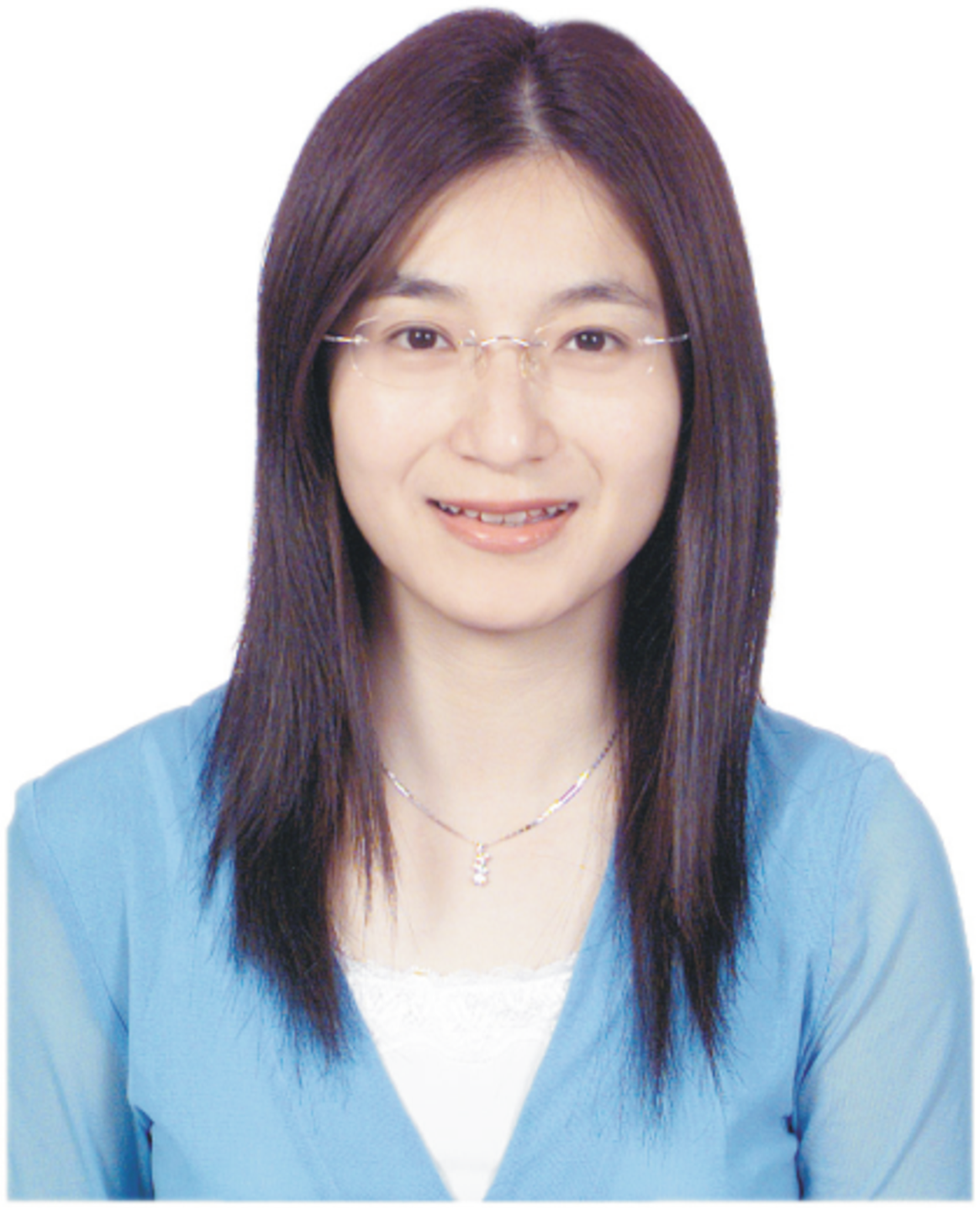}}]
{Wai M.~Tam} received the B.Sc. degree in
electronics and information systems from Jinan University, China,
and the M.Phil. and Ph. D. degree in electronic and information engineering from
The Hong Kong Polytechnic University, Hong Kong.\\
\hspace*{3ex}She is currently a Research Fellow at the Department of Electronic and
Information Engineering, The Hong Kong Polytechnic University, Hong Kong.
Her research interests include channel coding, mobile cellular systems, complex networks and chaos-based digital communications.
\end{IEEEbiography}

\end{document}